 \documentclass[final]{raa06}           %  For uploading to arXiv and final 
\usepackage{graphicx,times}             %for PS/EPS graphics inclusion, new
\usepackage{natbib}
\usepackage{amssymb,amsmath}
\bibpunct{(}{)}{;}{a}{}{,}
\usepackage[colorlinks=true, citecolor=blue]{hyperref}%
\usepackage{graphicx,kantlipsum,setspace}
\usepackage{caption}
\usepackage{graphics,epsf}
\usepackage{amsmath}                % American Mathematical Society package
\usepackage{amsfonts}               % American Mathematical Society fonts
\usepackage{amssymb}                % American Mathematical Society symbol
\usepackage{epsfig}                 % EPS figures
\usepackage{appendix}
\usepackage{graphicx}
\usepackage{float}
\usepackage{color}
\usepackage{multirow}
\usepackage{colortbl}
\usepackage[para,online,flushleft]{threeparttable}
\usepackage{xcolor}

\hypersetup{citecolor=blue, % color for \cite{...} links
            linkcolor=red, % color for \ref{...} links
            menucolor=blue, % color for Acrobat menu buttons
            urlcolor=blue}  % color for \url{...} links
\newcommand{\km}{{~\rm km}}

\newcommand{\keV}{{~\rm keV}}

\begin{document}

   \title{The vela supernova remnant: The unique morphological features of jittering jets
%\footnotetext{$*$ Supported by the National Natural Science Foundation of China.}
}
%   \subtitle{I. Place Your Subtitle Here}

   \volnopage{Vol.0 (20xx) No.0, 000--000}      %%preserved for Editor. DOn't remove!
   \setcounter{page}{1}          %%starting page, preserved for Editor. DOn't remove!

%\author[0000-0003-0375-8987]{Noam Soker}
%\author[0000-0002-9444-9460]{Dmitry Shishkin}

   \author{Noam Soker, Dmitry Shishkin
     % \inst{1}
    }

   \institute{Department of Physics, Technion, Haifa, 3200003, Israel;   {\it    soker@physics.technion.ac.il; s.dmitry@campus.technion.ac.il}\\
%        \and
%             Full institute address for the third author\\
\vs\no
   {\small Received~~20xx month day; accepted~~20xx~~month day}}

\abstract{We identify an S-shaped main-jet axis in the Vela core-collapse supernova (CCSN) remnant (CCSNR) that we attribute to a pair of precessing jets, one of the tens of pairs of jets that exploded the progenitor of Vela according to the jittering jets explosion mechanism (JJEM). A main-jet axis is a symmetry axis across the CCSNR and through the center. We identify the S-shaped main-jet axis by the high abundance of ejecta elements, oxygen, neon, and magnesium. We bring the number of identified pairs of clumps and ears in Vela to seven, two pairs shaped by the pair of precessing jets that formed the main-jet axis. The pairs and the main-jet axis form the point-symmetric wind-rose structure of Vela. The other five pairs of clumps/ears do not have signatures near the center, only on two opposite sides of the CCSNR.  We discuss different possible jet-less shaping mechanisms to form such a point-symmetric morphology and dismiss these processes because they cannot explain the point-symmetric morphology of Vela, the S-shaped high ejecta abundance pattern, and the enormous energy to shape the S-shaped structure. Our findings strongly support the JJEM and further severely challenge the neutrino-driven explosion mechanism. 
}

 \authorrunning{N. Soker, D. Shishkin}            
\titlerunning{The Vela SNR: jittering jets}  
   
      \maketitle

% ================================================
\section{Introduction}
\label{sec:Introduction}
% ================================================

Recent studies discuss two alternative theoretical explosion mechanisms of core-collapse supernovae (CCSNe), the delayed neutrino-driven mechanism and the jittering jets explosion mechanism (JJEM; for a most recent review, see \citealt{Soker2024Rev}). Recent studies of the neutrino-driven mechanism focus on three-dimensional simulations, starting with the pre-collapse stellar core to seconds after the revival of the stalled shock at $\simeq 100 \km$ from the newly born neutron star (NS; e.g., \citealt{Burrowsetal2024, JankaKresse2024, Muler2024, Mulleretal2024, vanBaaletal2024, WangBurrows2024, Nakamuraetal2025}; some studies consider hadron-quark phase transition in the frame of the neutrino-driven explosion; e.g., \citealt{Huangetal2024}). The magnetorotational explosion, which occurs when the progenitor core rotates rapidly and possesses strong magnetic fields (e.g., \citealt{Shibagakietal2024, ZhaMullerPowell2024}), is included under the neutrino-driven as it still attributes most CCSNe to the neutrino-driven mechanism and only rare cases to jet-driven with a fixed axis. 

On the other hand, recent studies of the JJEM focus on finding signatures of jittering jets in CCSN remnants (CCSNRs; e.g., \citealt{ShishkinKayeSoker2024, Soker2024key, Soker2024NA1987A, Soker2024CF, Soker2024CounterJet, Soker2024PNSN, BearShishkinSoker2025, BearSoker2025, ShishkinSoker2025}). Last year's findings of such signatures in several CCSNRs and other expected signatures of the JJEM have led to a big step forward in establishing the JJEM as the main, or even sole, explosion mechanism of CCSNe (reviews by  \citealt{Soker2024Rev, Soker2024UnivRev}). Neutrino heating plays a role in the JJEM but not the primary role. Namely, neutrino heating can help the launching of the jets from the intermittent accretion disks (or belts) around the newly born NS (but magnetic fields are also needed), and neutrino can boost the jet energy after they were launched \citep{Soker2022nu}.

In the JJEM, pairs of jets with varying directions that the newly born NS, or black hole in some cases, launches explode the star (e.g., \citealt{Soker2010, PapishSoker2011, PapishSoker2014Planar, GilkisSoker2014, GilkisSoker2016, Soker2020RAA, Soker2022Rev}). The source of the stochastic angular momentum of the gas that the NS accretes is the pre-collapse convective angular momentum fluctuations in the core (e.g., \citealt{ShishkinSoker2021, ShishkinSoker2023}) that instabilities above the newly born NS amplify, mainly instability modes of the spiral standing accretion shock instability (SASI; e.g.,  \citealt{Andresenetal2019, Walketal2020, Nagakuraetal2021, Shibagakietal2021}, for spiral-SASI). Envelope convection can be the seeds of the angular momentum fluctuations in electron-capture supernovae (\citealt{WangShishkinSoker2024}). If, for some reason, the accretion of core material does not lead to the explosion and a black hole is formed, then the envelope convection can also be the seed of the angular momentum fluctuations  (\citealt{Quataertetal2019,  AntoniQuataert2022, AntoniQuataert2023}). 

According to the JJEM, a black hole is formed when, because of a rapidly rotating pre-collapse core, the central newly born NS (before collapsing to a black hole) launches the exploding jets along a fixed axis that implies inefficient jet feedback mechanism (e.g., \citealt{Soker2023gap}). A large amount of accreted mass can result in a super-energetic CCSN (e.g., \citealt{Gilkisetal2016}). There is a small jittering around the fixed axis. 

The JJEM has several to a few tens jet-launching episodes (e.g., \citealt{Soker2025Learing}). The stellar core material that did not collapse to the newly born NS chocks most of these jets and acquires their energy; this leads to the core explosion. From that time, the explosion is similar in many aspects, but not all, to the neutrino-driven explosion mechanism; e.g., there are many instabilities, and the NS acquires a kick. However, the later jets expand more freely and can leave imprints on the ejecta; these jets are still part of the exploding jets and are not post-explosion jets. Each late jet pair can leave two opposite (relative to the center of the explosion) morphological features. If two or more pairs of jets are along different axes, the outcome is a point-symmetric morphology. The point-symmetric morphology of the Vela CCSNR is compatible with the JJEM (e.g., \citealt{Soker2023SNRclass, Soker2024CF}).

In many cases, the two opposite jets in a pair will not be equal in their power and opening angle because of the short-lived launching accretion disk \citep{Soker2024CounterJet}, {{{{ and might impart a kick to the NS }}}} \citep{BearShishkinSoker2025}. 
Unequal opposite jets in a pair also exist in young stellar objects, e.g., \cite{Murphyetal2024}. 
   
The JJEM allows for the last pair or two to be long-lived and hence have a sizeable morphological impact in forming a morphological feature from one side through the center to the other side of the CCSNR \citep{Soker2024key}. Such are the main jet-axis of SN 1987a, the keyhole structure \citep{Soker2024key}, the main jet-axis of SNR 0540-69.3 \citep{Soker2022SNR0540}, and the main jet-axis, the S-shaped hose of the Cygnus Loop \citep{ShishkinKayeSoker2024}. In this study, we reveal the main-jet axis of the Vela SNR (section \ref{sec:Vela}) and argue that only jets can shape it. The point-symmetric morphology of the Vela CCSNR leaves only the JJEM as a viable explanation for its explosion and shaping. We add this finding to earlier findings of other CCSNRs to compare the JJEM with possible alternatives for these morphological features; we find that only the JJEM can account for all morphological properties (Section \ref{sec:PointSymmetry}). 
We summarize this study in Section \ref{sec:Summary}.

% ================================================
\section{Identifying the main axis of the Vela CCSNR}
\label{sec:Vela}
% ================================================
% ================================================
\subsection{The point-symmetric wind-rose of Vela}
\label{subsec:Wind-rose}
% =============================================

\begin{figure*}[t]
\begin{center}
\includegraphics[trim=2cm 0.5cm 2cm 1.5cm, clip, width=\textwidth]{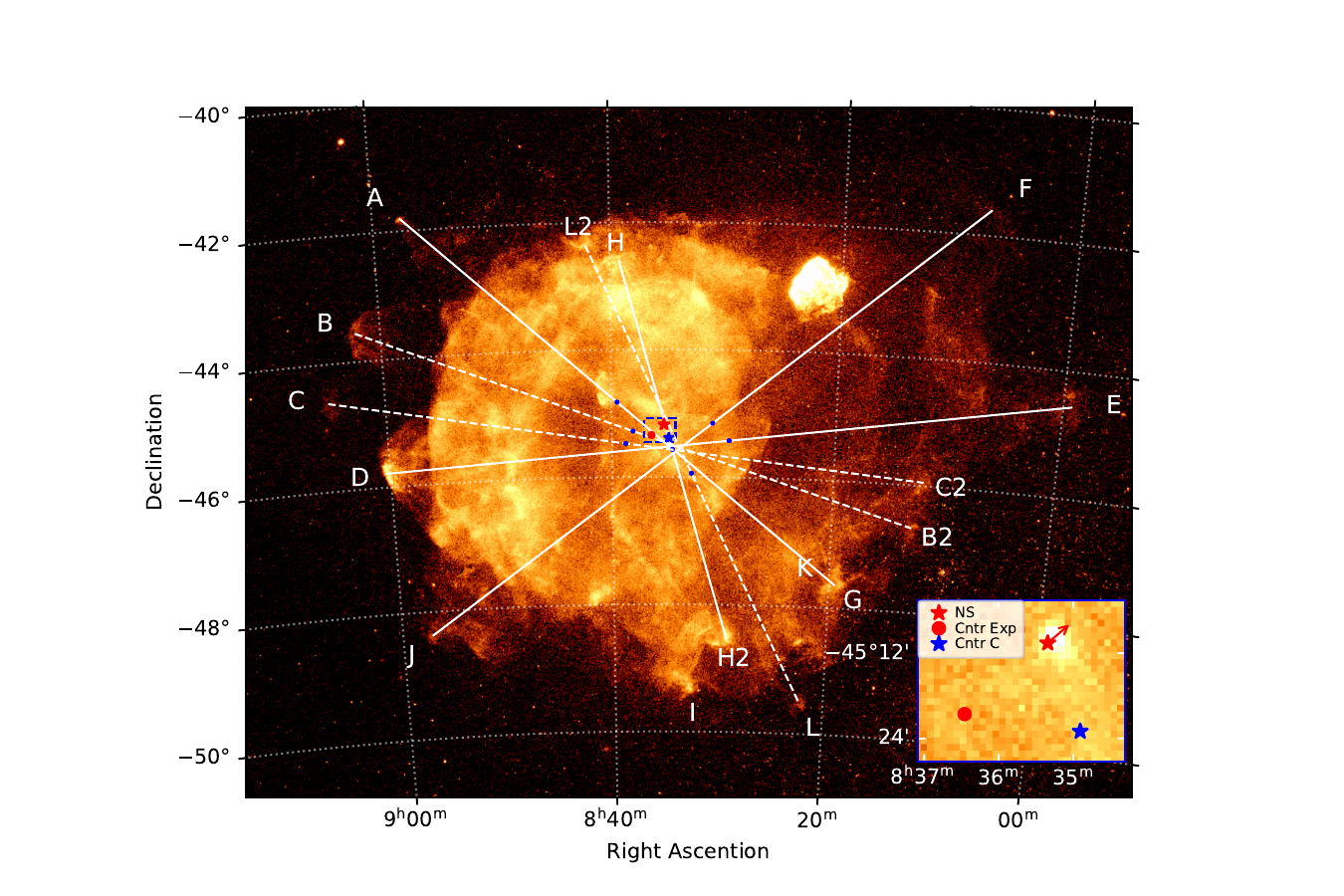} %[trim=left lower right upper]
\caption{An x-ray counts image from the eROSITA DR1 (log scale, $0.2-2.3~\rm{keV}$). Solid lines connect previously identified pairs of clumps, and dashed lines are pairs we identify in the present study; these are the symmetry axes of the point-symmetric wind-rose. We mark the mid-point of each symmetry axis with a blue dot. We mark the average location of these centers with a blue asterisk. In an inset on the bottom right ($29.2^{\prime} \times 22.7^{\prime}$), we present the inner part of the Vela SNR, including the NS location (\citealt{Kochanek2022}; red asterisk), its projected movement direction (red arrow), and the presumed origin at explosion (\citealt{Kochanek2022, Dodson_etal_2003}; red dot).}
\label{fig:WindRose}
\end{center}
\end{figure*}

Earlier studies of Vela identified several clumps of Vela; these are clumps A-L in Figure \ref{fig:WindRose}. \cite{Aschenbachetal1995} marked clumps A-F, \cite{Garciaetal2017} added clumps G and drew the line AG, \cite{Sapienzaetal2021} added more clumps' labeling, and \cite{Mayeretal2023} identified the possible clump L. \cite{Sapienzaetal2021} argued that clumps K and G are counter to clump A and are jet-like structure from the explosion process. \cite{Soker2024CF} identified clump H2 from X-ray images by \cite{Mayeretal2023} and extended the point-symmetric wind-rose of Vela from \cite{Soker2023SNRclass} to include the symmetry lines (axes) AG, DE, FJ, and HH2. 
% FFFFFFFFFFFFFFFFFFFFFFFFFFFFFFFFFFFFFFFFFF
% FFFFFFFFFFFFFFFFFFFFFFFFFFFFFFFFFFFFFFFFFF

The high Si abundance of clumps A \citep{KatsudaTsunemi2006}, G, and K  \citep{Garciaetal2017} implies that they originate deep inside Vela's progenitor's core.  \cite{KatsudaTsunemi2005} found clump D to have ONeMg overabundance, indicating its origin from near the remnant's center, as previously \cite{Sankritetal2003} suggested. \cite{GrichenerSoker2017ears} took ears D and E to compose the main jet-axis of Vela and, {{{{ based on the relative volume of the ears, }}}} estimated the total energy of the two jets that inflated these ears to be $\approx 1 \%$ of the Vela explosion energy, which is very low energy.  We identify a new main-jet axis (section \ref{subsec:MainAxis}), which, together with the other jet pairs {{{{ have a total volume $\approx 20$ times as large as that of ears D and E. The identified pairs, therefore, }}}}  can bring the total energy of the shaping jets to be $\approx 20 \%$ of the explosion energy. According to the JJEM, the rest of the explosion energy is due to earlier jets that did not leave an imprint on the point symmetric morphology because they exploded the core of the stellar progenitor. 

Although clumps B2 and C2, which we define in figure \ref{fig:WindRose} are small, they are ($i$) prominent in their surroundings, and ($ii$) not smaller than clumps A and L and not much smaller than clumps G that were identified in the past as clumps. The last property is shared by clump L2, which we also identify in Figure \ref{fig:WindRose}. When we connect B2 to B, C2 to C, and L2 to L (three dashed lines in Figure \ref{fig:WindRose}), the three lines cross the center defined by the other five symmetry lines (up to the uncertainty in the centers of the clumps at the ends of the lines). In section \ref{subsec:MainAxis}, we argue that pairs LL2 and HH2 belong to one pair of precessing jets. 

Three crucial comments regarding the seven pairs of clumps composing Vela's point-symmetric wind-rose are in place here. (1) In the JJEM, the two jets in each pair of jets are expected to be unequal in their opening angle and power because the intermittent accretion disk that launches the pair of jets has no time to fully relax \citep{Soker2024CounterJet}.  Therefore, the centers of the symmetry lines, marked by blue dots, might miss the lines' cross-point, and not all lines will cross at precisely the same point. (2) Not each pair of clumps are necessarily the heads of two opposite jets. Dense clumps might be formed in compressed zones between jet-inflated bubbles. This is observed in the hot gas of clusters of galaxies, e.g., Abell 2597 \citep{Tremblayetal2018}, and was found in the hydrodynamical simulations of the early phase of the JJEM \citep{PapishSoker2014Planar}.  (3) The distance of the average centers of the lines (blue asterisk) from the NS location (red asterisk) and the distances of the cross points of the symmetry axes with each other from the NS location are similar to the distance the NS has moved from the explosion site to its present location. Considering the unequal jets in a pair (point 1), the center of the seven symmetry axes is sufficiently close to the present location of the NS and to the location of the NS at the explosion to be associated with the explosion. Namely, pairs of jets exploded the progenitor of the Vela SNR. 

We turn to identify Vela's main-jet axis. 

% ================================================
\subsection{The main-jet axis of Vela}
\label{subsec:MainAxis}
% ================================================

In this section, we use the new results of the eROSITA X-ray telescope to identify Vela's main-jet axis. A main-jet axis is one across the diameter of the SNR, i.e., with a structure inside the main SNR shell in addition to the SNR outer zones. For example, in the most prominent pair of ears, the ears (clumps) D and E have prominent structures in the outer zones of the SNR. However, near the center, no signatures are related to the symmetry axis connecting ears D and E.  

\begin{figure}[t]
\begin{center}
\includegraphics[trim=0cm 0cm 0cm 0cm, clip, width=0.5\textwidth]{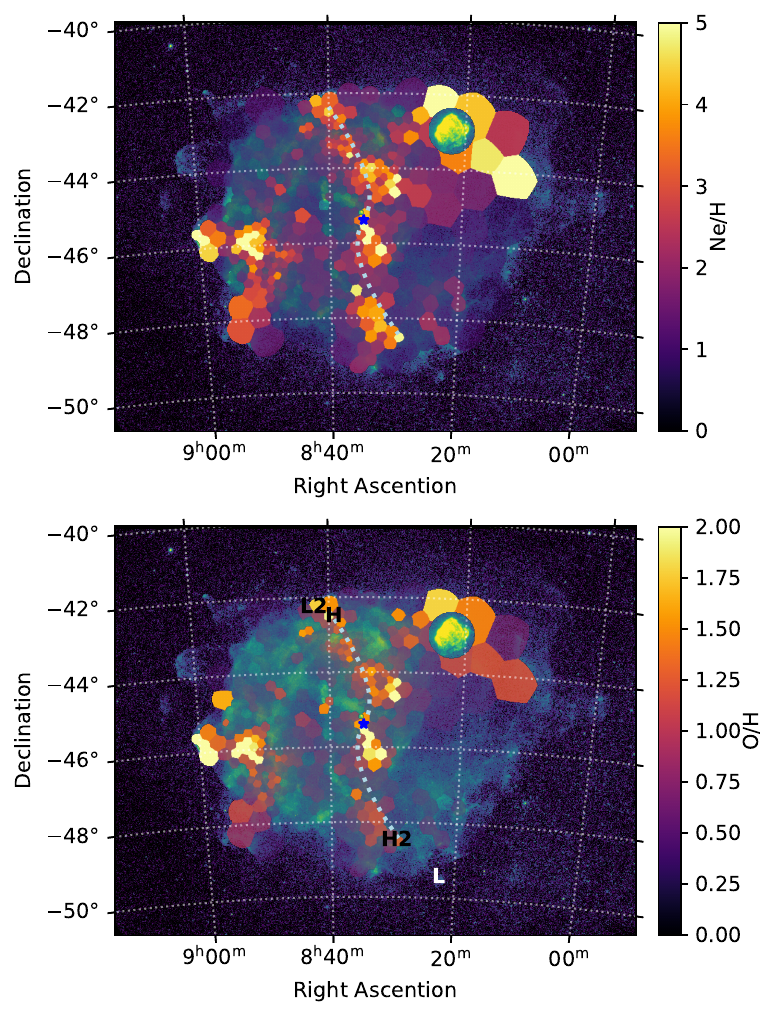} %[trim=left lower right upper]
\caption{Relative abundances map of Ne/H (upper panel) and O/H (lower panel) adapted from \cite{Mayeretal2023} (semi-transparent, color bar on the right). We overlay the abundance maps on the X-ray counts image from Figure~\ref{fig:WindRose} (`viridis' color map) to guide the eye. An enhanced neon abundance S-shaped zone goes from north to south through the center; based on this zone, we mark an S-shaped dotted teal line along the main-jet axis. Mg/H (not shown) abundances indicate a similar feature. A blue asterisk marks the {{{{average location of the centers of clump pairs}}}}, i.e., the center of the point-symmetric structure (Figure~\ref{fig:WindRose}). The plotted S-shaped line has a 180-degree symmetry around this center. We denote the HH2 and LL2 clumps on the lower panel for reference.} 
\label{fig:serp}
\end{center}
\end{figure}

\cite{Mayeretal2023} performed an extensive study of the Vela SNR in the X-ray, using the eROSITA DR1 data. Their results include some abundance distributions in the SNR as contrived from a spectral model fitted to the X-ray data. They identify enhanced abundances (relative to solar ratios) of oxygen, neon, and magnesium along a zone extending north to south and through the remnant's center. In Figure~\ref{fig:serp}, we present the distribution of neon and oxygen abundance. The region of enhanced neon and oxygen abundance through the center has an S-shaped morphology (and so is magnesium in the maps that \citealt{Mayeretal2023} present). The S-shaped structure, which we identify as the main-jet axis and draw by a dotted line in Figure \ref{fig:serp}, includes on its northern end clumps H and L2, and its southern end clumps H2 with L further out. We attribute the pairs HH2 and LL2 to the same pair of jets, a precessing jet pair.  
% FFFFFFFFFFFFFFFFFFFFFFFFFFFFFFFFFFFFFFFFFF
% FFFFFFFFFFFFFFFFFFFFFFFFFFFFFFFFFFFFFFFFFF

To the immediate south of where we denote our center (blue asterisk in Figure~\ref{fig:serp}) is an enhanced abundance feature that slightly deviates from our denoted S-shape. This feature coincides with a ``cocoon'' region to the south of the pulsar wind nebula (PWN), which \cite{Slane_etal_2018} have attributed to a result of the reverse shock interacting with the PWN - hence, we do not expect it to be consistent with other features that we associate to the explosion. Indeed, as we emphasize in this study, the PWN, alongside other processes, smear the point-symmetrical morphology. 

\begin{figure}[]
\begin{center}
\includegraphics[trim=3.5cm 2.5cm 0cm 3.3cm, clip, width=0.6\textwidth]{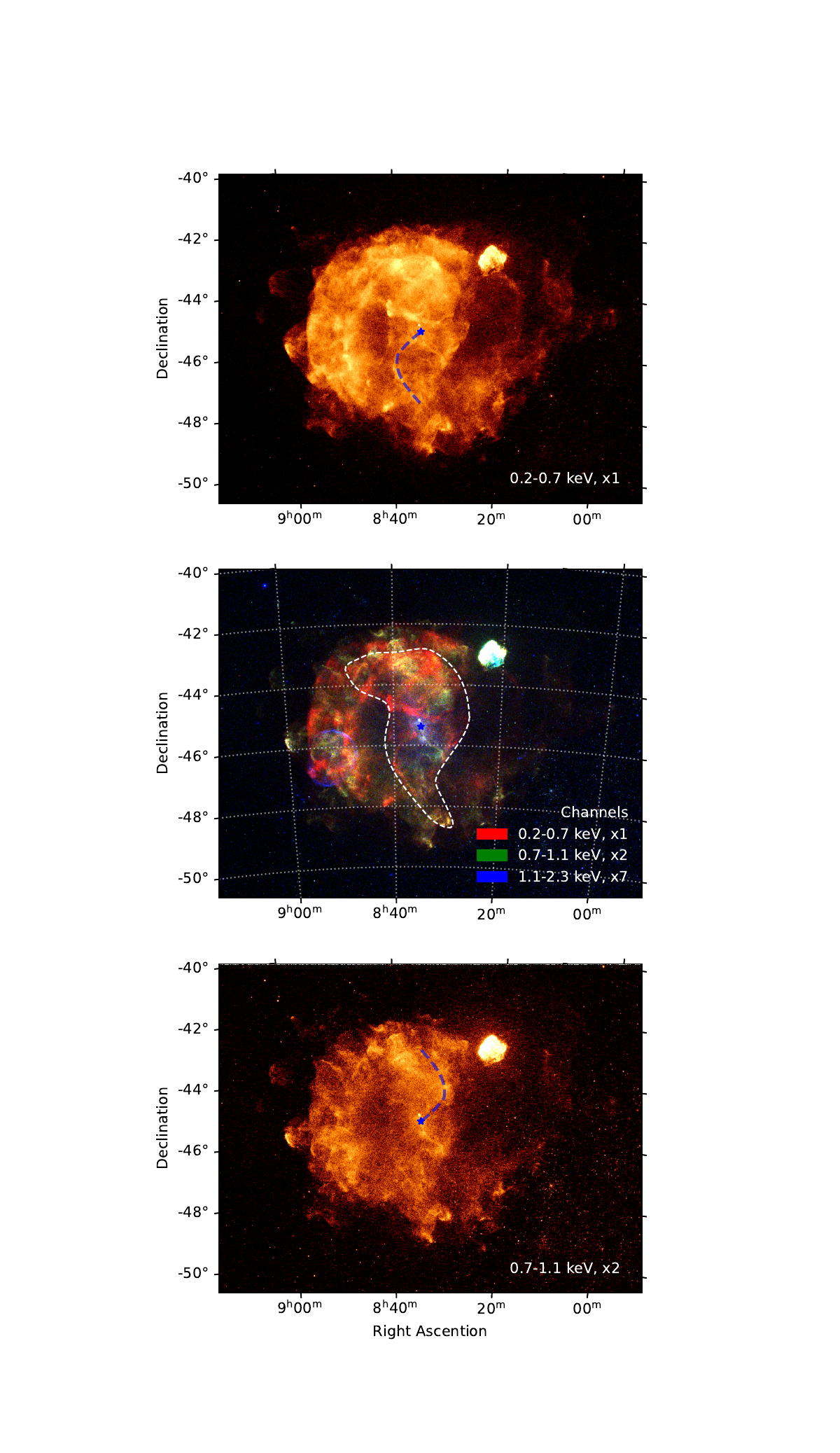} %[trim=left lower right upper]
\caption{X-ray count images of the Vela SNR in different energy bands. \textbf{Top panel:} log scaled X-ray image in the $0.2-0.7 \keV$ range. We denote a sharp filamentary structure most apparent in this range with a dashed blue line. \textbf{Middle panel:} a composite RGB X-ray image, linear scale, of the three energy bands, with a dashed white line denoting the central inner north-south S-shaped structure (see text). The three channels are balanced to account for the lesser counts of the less bright energy ranges. \textbf{Bottom panel:} log scaled X-ray image in the $0.7-1.1 \keV$ range. We rotate the dashed blue curve from the top panel by 180 degrees to point out its alignment with features on the northern part, most visible in this range. We denote the center of the point-symmetric structure in all panels with a blue asterisk.}
\label{fig:traced}
\end{center}
\end{figure}

In Figure \ref{fig:traced}, we present the Vela SNR X-ray image separated by energy ranges, soft, $0.2-0.7 \keV$ (upper panel), combined, including the higher range in which the SNR is still visible, $1.1-2.3 \keV$ (middle panel), and  $0.7-1.1 \keV$ (lower panel). In the upper and lower panels of Figure \ref{fig:traced}, we draw two symmetric sides of an S-shaped axis (dashed lines), different from the one that we draw in Figure \ref{fig:serp}. 
{{{{ These two sides are less secure than those in Figure \ref{fig:serp}. The southern line is along a thin X-ray filament that might not be a jet-axis. The northern one is parallel and close to the northwest side of the dashed-white line we draw in the middle panel. This more bent (than in Figure \ref{fig:serp}) S-shape structure might not signify the tracks of the two jets. Nonetheless, }}}} 
the difference between the two S-shaped axes conveys the uncertainty in the exact center of the north-south S-shaped structure.  
% FFFFFFFFFFFFFFFFFFFFFFFFFFFFFFFFFFFFFFFFFF
% FFFFFFFFFFFFFFFFFFFFFFFFFFFFFFFFFFFFFFFFFF

We identify another structure that we attribute to the pairs of precessing jets that formed the main-jet axis. The images in Figure \ref{fig:traced} reveal a sharp boundary between an inner elongated bright S-shaped structure and a fainter outer structure. We mark this boundary in the middle panel of Figure \ref{fig:traced}. In some segments of the closed boundary, the jump in brightness is clear, and in some, it is less clear. This boundary is around the S-shaped main-jet axis. We suggest that this boundary-enclosed material was shaped by the energetic jets of the main-jet axis. 

\begin{figure*}[]
\begin{center}
\includegraphics[trim=4cm 1cm 4cm 2.5cm, clip, width=\textwidth]{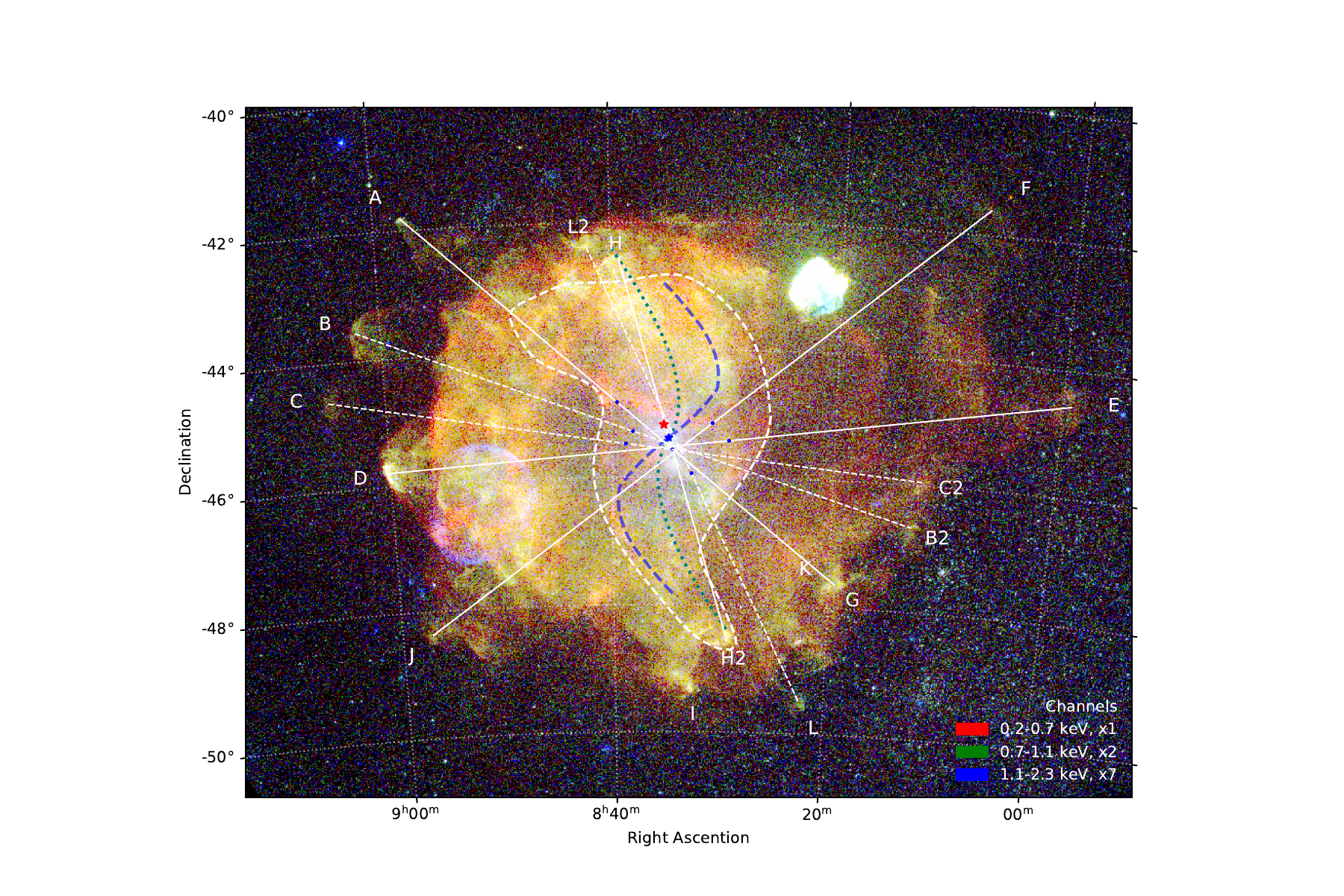} %[trim=left lower right upper]
\caption{RGB-colored log scaled X-ray counts image of the Vela SNR in different energy bands. Ranges and scaling are as in the middle panel of Figure~\ref{fig:traced}. We mark all the features we identified. Clumps are denoted with a letter in an away-from-center direction. Solid white lines connect clumps established as pairs in previous studies, while dashed lines denote pairs we identify here (Figure~\ref{fig:WindRose}). The dotted teal line is from Figure~\ref{fig:serp} and marks the general location of the central enhanced abundances feature identified by \cite{Mayeretal2023}. The dashed blue line marks a sharp filamentary structure passing through the center as we drew in the top and bottom panels of Figure~\ref{fig:traced}. The dashed white curve marks the boundary of the central inner elongated structure (middle panel of Figure~\ref{fig:traced}). We mark the center of the point-symmetric structure with a blue asterisk and the neutron star location with a red asterisk. The bright white structure in the northwest is the Puppis A SNR, and the faint round blue structure in the southeast is the Vela Jr SNR.}
\label{fig:overlayed}
\end{center}
\end{figure*}

In Figure \ref{fig:overlayed}, we draw the point-symmetric wind-rose from Figure \ref{fig:WindRose}, the dotted S-shaped line from Figure \ref{fig:serp}, and the dashed-blue S-shaped line and the dashed-white boundary from Figure \ref{fig:traced} on the same X-ray image of Vela. The point symmetric structure appears here in its full glory. We emphasize that the two S-shaped lines (dotted and dashed) are not two jet pairs but one precessing pair of jets with uncertain locations; this is the main-jet axis of Vela. The HH2 axis and LL2 axis do not represent separate jet pairs but rather clumps that belong to the same precessing jet pair. 
% FFFFFFFFFFFFFFFFFFFFFFFFFFFFFFFFFFFFFFFFFF
% FFFFFFFFFFFFFFFFFFFFFFFFFFFFFFFFFFFFFFFFFF

Near the center, the direction of the S-shaped structure that we draw with the dashed blue line in figures \ref{fig:traced} and \ref{fig:overlayed} is parallel to the jets' direction of the pulsar of the Vela SNR, PSR~B0833-45 (e.g., \citealt{Helfandetal2001, Fateevaetal2023}). Namely, to the accuracy in determining the S-shaped line, e.g., compare the dashed and dotted S-shaped lines in Figure \ref{fig:overlayed}, near the center, the S-shaped segment is more or less parallel to the spin of the pulsar. However, the symmetry axes HH2 and LL2 are almost perpendicular to the spin direction. 

We also note that the PWN covers a small region of the sky, only $\simeq 8^{\prime \prime}$ across (e.g., \citealt{Helfandetal2001, Fateevaetal2023}). Therefore, the shaping of the S-shaped structure cannot be due to the PWN.

% ================================================
\section{Observational evidence for point-symmetric exploding jets}
\label{sec:PointSymmetry}
% ================================================

According to the JJEM, energetic jets, as expected to explode the star, can account for the properties of the point-symmetric structures of CCSNRs, including pairs of opposite clumps, filaments, ears, lobes, and nozzles. 
Other processes can also contribute to the shaping of CCSNRs but cannot explain the majority of observed point-symmetric morphologies. In Table \ref{Tab:Table1}, we list four shaping processes and whether they can or cannot account for specific morphologies of some CCSNRs. Although the interstellar medium (ISM) also influences the morphology of CCSNRs (e.g., \citealt{Sofue2024} for a very recent study), it cannot explain the basic morphological features we are studying here, in particular point symmetry. The ISM does poorer than the CSM in accounting for the CCSNR properties we are interested in. Only the JJEM can account for all the morphological structures we examine in the Table.  
% TTTTTTTTTTTTTTTTTTTTTTTTTTTTTTTTTTTTTTTTTTTTTTTTTTTTTT
\begin{table*}
%\tiny
%\scriptsize
%\footnotesize
\begin{center}
  \caption{Shaping processes of CCSNRs against observations}
    \begin{tabular}{| p{3.4cm} | p{3.2cm} | p{3.2cm}| p{3.2cm}| p{3.2cm} | }
\hline  % ----------------------------
\textbf{{}}  & \textbf{Exploding jets} & \textbf{Post-explosion jets} & \textbf{CSM} &  \textbf{Instabilities}  \\
\hline  % ----------------------------
The Si/Mg jet of CasA and ejecta-rich material of clumps in Vela$^{[{\rm C01}]}$. 
    &  The jet interacted with the layer that nucleosynthesis Si/Mg.  
    & \textcolor{red}{- Interaction with low-density ejecta will not synthesize Si/Mg.}  
    & \textcolor{red}{- Cannot explain.}  
    & \textcolor{red}{- Instabilities have all synthesised metals in the fingers$^{[{\rm C02}]}$}. \\
\hline  % ----------------------------
The [Ar \textsc{ii}] map of CasA$^{[{\rm C03]}}$ is point symmetric$^{\rm [C06]}$. 
    & \textcolor{green}{$\bigoplus$} Expected in the JJEM$^{\rm [C06]}$. 
    & Possible if the fallback gas has varying angular momentum axis. 
    & \textcolor{red}{- The argon is inside an outer ejecta, so the CSM has no influence.} 
    & \textcolor{red}{- Do not give consistent opposite pairs.} \\  
   \hline  % ----------------------------
CasA's main ear-pair carry $\simeq 10 \% E_{\rm exp}$$^{\rm [C04]}$. JWST$^{\rm [C05]}$  reveal more point-symmetric pairs$^{\rm [C06]}$, 
    & \textcolor{green}{$\bigoplus$} The energies of pairs are in the expected range of the JJEM$^{\rm [C07]}$. 
    & \textcolor{red}{- The energy is much larger than expected in post-explosion jets unrelated to 
          the explosion.} 
    & Massive CSM can explain the energetics. \textcolor{red}{- The JWST delicate point-symmetry is hard to explain. } 
    & Can explain the energetics. \textcolor{red}{- The JWST delicate point-symmetry is hard to explain. } \\  
   \hline  % ----------------------------
The rim-nozzle asymmetry of the `keyhole' of SN 1987A ejecta.   
   & \textcolor{green}{$\bigoplus$} Expected in some CCSNRs, as in many other jet-shaped astrophysical objects$^{{\rm [C08]}}$.  
   & \textcolor{red}{- Structure too large to be explained by jets unrelated to the explosion.} 
   & \textcolor{red}{- Did not reach yet the inner CSM ring.} 
   & It is possible, but two fingers are unlikely to be exactly opposite.  \\  
   \hline  % ----------------------------
Point-symmetric clumps in the ejecta of SN 1987A$^{\rm [C09]}$.
   & \textcolor{green}{$\bigoplus$} Expected in some CCSNRs$^{\rm [C10]}$. 
   & Possible if the fallback gas has varying angular momentum axis. 
   & \textcolor{red}{- Did not reach yet the inner CSM ring.} 
   &  \textcolor{red}{- Do not give consistent opposite pairs.} \\  
   \hline  % ----------------------------
SNR 0540-69.3 has a main jet axis in HST images and
Doppler maps show point symmetry$^{\rm [C11]}$.  
   & \textcolor{green}{$\bigoplus$} Main jet-axis + rings like jet-shaped galaxy cluster Cygnus A$^{\rm [C10]}$. Point symmetry is expected.
   & \textcolor{red}{- Structure too large to be explained by jets unrelated to the explosion.}  
   & \textcolor{red}{- Point-symmetry also exists near the remnant's center. }
   & \textcolor{red}{- Four pairs of point-symmetric clumps are highly unlikely by stochastic instabilities.} \\  
\hline  % ----------------------------
Three pairs of unequal opposite ears in SNR N63A. Bright rims on front of ears.   
& \textcolor{green}{$\bigoplus$} Unequal opposite jets compatible with the JJEM$^{\rm [C12]}$. Ears' bright rim as in other jet-inflated ears$^{\rm [C13]}$. 
   & Possible if the fallback gas has varying angular momentum axis.
   & \textcolor{red}{- Imprints of ears in inner regions + unequal opposite ears are hard to explain.}  
   & \textcolor{red}{- Unlikely to inflate opposite ears; some are wide. } \\  
\hline  % ----------------------------
Vela SNR has seven pairs of opposite ears and clumps $^{\rm [C0; C14]}$. 
   & \textcolor{green}{$\bigoplus$} The interaction of jet-inflated bubbles in the core can form dense clumps$^{\rm [C15]}$. 
   & Possible if the fallback gas has varying angular momentum axis. 
   & \textcolor{red}{- Fast clumps are incompatible with shaping by the CSM.} 
   & \textcolor{red}{- Stochastic instabilities unlikely to eject seven pairs of opposite clumps. } \\  
\hline  % ----------------------------
Vela shows an O/Ne/Mg-rich S-shaped main axis extending from one side to the other through the center$^{\rm [C16]}$. 
   & \textcolor{green}{$\bigoplus$} A main axis of CCSNRs is explained by the JJEM$^{\rm [C17]}$. A precessing jet pair explains the S shape. 
   & \textcolor{red}{- O/Ne/Mg-rich material should be ejected at explosion. The structure is too large for jets unrelated to the explosion. } 
   & \textcolor{red}{- The composition and extension to the center are not compatible with CSM interaction. }
   & \textcolor{red}{- The S-shaped structure is incompatible with instabilities, even if two instability fingers are on opposite sides. } \\  
\hline  % ----------------------------
The large-scale elongated structure of SNR G321.3-3.9 $^{\rm [C18]}$. 
& \textcolor{green}{$\bigoplus$} The JJEM accounts for this by two main jet-launching episodes with close angles$^{\rm [C18]}$. 
& \textcolor{red}{- The elongated structure is too large for post-explosion jets that carry $\ll E_{\rm exp}$.} 
& \textcolor{red}{- The structure extends outward from inner regions, where CSM has no influence.} 
& \textcolor{red}{- The structure is too coherent for instabilities. Instabilities perturb the coherent structure. } \\  
%\hline  % ----------------------------
\hline  % ----------------------------
     \end{tabular}
  \label{Tab:Table1}\\
\end{center}
\begin{flushleft}
\small 
Note: The green plus sign indicates that the explanation for the observation exists in the literature. 
The red color indicates the process cannot or has problems explaining the observation. 
\newline
Abbreviation: CasA: Cassiopeia A; JJEM: jittering jets explosion mechanism; $E_{\rm exp}$: The explosion energy, including the ejecta energy plus radiation.  
\newline
Comments: 
C0: this study
C01: \cite{Grefenstetteetal2017} for Cassiopeia A and \cite{Mayeretal2023} for the Vela SNR;
C02: \cite{Wongwathanaratetal2015};
C03: \cite{DeLaneyetal2010};
C04: \cite{GrichenerSoker2017}
C05: \cite{Milisavljevicetal2024}; 
C06: \cite{BearSoker2025}; 
C07: \citep{Soker2024key}; 
C08: other astrophysical objects showing rim-nozzle asymmetry include planetary nebulae and hot gas in clusters of galaxies \citep{Soker2024key, Soker2024CF, Soker2024PNSN};
C09: \cite{Soker2024NA1987A};
C10: \cite{Soker2024CF}; 
C11: \cite{Soker2022SNR0540}, where HST images of SNR 0540-69.3 are from \cite{Morseetal2006} and Doppler-shift maps from \cite{Larssonetal2021};
C12: \cite{Soker2024CounterJet}; 
C13: The rims of the ears in SNR N63A are similar to rims in jet-inflated bubbles in clusters of galaxies \citep{Soker2024CF} and ears/bubbles/lobes in planetary nebulae \citep{Soker2024PNSN};
C14: \cite{Soker2024CF} identified four pairs. 
C15: \cite{PapishSoker2014Planar};
C16: \cite{Mayeretal2023} reveal the O/Ne/Mg structure, and in this study (section \ref{subsec:MainAxis}), we reveal the S-shaped main-jet axis of the Vela SNR;  
C17: \cite{Soker2024key}; 
C18: \cite{ShishkinSoker2025}.

\end{flushleft}
\end{table*}
% TTTTTTTTTTTTTTTTTTTTTTTTTTTTTTTTTTTTTTTTTTTTTTTTTTTTTTTTTTTTTTTT

We did not include shaping by a possible PWN because, in most CCSNRs we study, there is no indication of a PWN (e.g., SN 1987A). Also, in general, PWN power is insufficient to explain the shaping of large structures along the polar directions. Instead, the PWN's shocked material fills the CCSNR volume.  

According to the neutrino-driven explosion mechanism, post-explosion jets might shape CCSNRs, like Cassiopeia A (e.g., \citealt{Orlandoetal2021}). According to Table \ref{Tab:Table1}, shaping by post-explosion jets can take place in some cases (only in some cases) if the axes of pairs of jets change between jet launching episodes and the jets are energetic. This raises the question of why earlier jets could not be launched either; these early jets then explode the star. In some cases, the post-explosion jets cannot account for the observed properties. Only exploding jets with varying directions can account for all observed morphologies. 
Indeed, \cite{Jankaetal2022spin} discuss the formation of pairs of ears by fallback material (see also \citealt{Muller2023spin}), but comment that this outflow is not expected to be energetic. 
\cite{AkashiSoker2022} also simulated post-explosion jets that can only shape the very inner zones of the ejecta; this also holds for jets launched by an NS companion (e.g., \citealt{AkashiSoker2020}). \cite{AkashiSoker2021} show that very late jets, launched weeks after the explosion, can power a peak in the light curve. However, these jets cannot form large-scale point-symmetric morphologies.  
The famous case of SNR W50 is observed to be shaped by precessing jets that the central binary system SS433 launches. This is a different category of processes that do not belong to this study because it involves a post-explosion active binary system.    

We emphasize that all the processes we list can take place. In particular, we expect the ISM (e.g.,  \citealt{Wuetal2019, YanLuetal2020, LuYanetal2021, MeyerMelianietal2024}), CSM (e.g., \citealt{Chiotellisetal2021, ChiotellisZapartasMeyer2024, Velazquezetal2023, Meyeretal2022, MeyerDetal2024}) and instabilities (e.g., \citealt{Wongwathanaratetal2015}) to play a role in shaping the majority of CCSNe (the CSM might influence only older CCSNRs, as the CSM of SN 1987A did not affect the inner massive ejecta yet).  Also, an NS natal kick might be similar (\citealt{BearSoker2023RNAAS}) to that in the neutrino-driven explosion mechanism (e.g., \citealt{Wongwathanaratetal2013kick}) and/or might result from an early asymmetrical jet pair, the kick by early asymmetrical pair (kick-BEAP) mechanism \citep{BearShishkinSoker2025}; the kick direction avoids small angles with the main-jet axis (e.g., \citealt{BearSoker2023}). 
However, instabilities and the ejecta-CSM interaction smear and dilute the point-symmetric morphology rather than produce it. 
Moreover, the hot ejecta itself expands and smears point-symmetrical features; hot ejecta results from the explosion itself, the decay of nickel (nickel bubbles; e.g., \citealt{MilisavljevicFesen2013}), and the reverse shock (e.g., \citealt{HwangLaming2012}).  
For these smearing processes, it is hard to identify point-symmetric morphological components in many cases. The NS kick, accompanied by asymmetrical mass ejection, adds to the smearing of point symmetry. The main properties of instabilities and ejecta-CSM interaction in the JJEM are similar to those in the neutrino-driven explosion mechanism.  The JJEM has jets that carry more energy than instabilities. Hence, the jets can form point-symmetric morphological structures. Instabilities, nonetheless, somewhat smear the point symmetry.

Other features further support the JJEM. 
\newline
\textit{Jittering in a plane.} 
The ejecta of Cassiopeia A is concentrated around one plane \citep{MilisavljevicFesen2013}.  \cite{PapishSoker2014Planar} speculated that the torus morphology of a tilted thick disc with multiple jets in Cassiopeia A, as observations find (e.g., \citealt{Willingaleetal2003, DeLaneyetal2010, MilisavljevicFesen2013}), results from the tendency of jittering jets in the JJEM to share a plane, up to the fluctuations that change jets' axes, and which can even avoid the planar jittering. \cite{BearSoker2025} confirmed this speculation by identifying the point symmetry of Cassiopeia A; because dense clumps, which are concentrated in this plane, are much brighter than their surroundings, there might be bias in emission from this plane (e.g., \citealt{HwangLaming2012}), implying the possible presence of massive ejecta also perpendicular to this plane.  A planar jittering might have also shaped SNR 0540-69.3; this requires further study. 
\newline
\textit{Main-jet axis.} This is the case with the `keyhole' structure of SN 1987A, the jet axis of SNR 0540-69.3, and the S-shaped hose that is the main jet-axis of the Cygnus loop \citep{ShishkinKayeSoker2024}. The explanation in the frame of the JJEM \citep{Soker2024key} is that at the end of the accretion process onto the newly born NS, the mass accretion decreases so that the time scale of the fluctuations of the angular momentum increases, allowing long-lived jet-launching episodes; in particular the last one. This last pair of jets might form the main-jet axis. Note that this axis need not be along the pre-collapse rotation axis of the core in case the pre-collapse rotation is slow. 
An interaction with a CSM shapes the outskirts of the SNR; it cannot form a main jet-axis with imprints in the remnant's center, as with the Vela SNR (Section \ref{subsec:MainAxis}). Based on the above, we dismiss the suggestion of \cite{Gvaramadze2000} that a CSM shaped the Vela SNR. 

% ================================================
\section{Summary}
\label{sec:Summary}
% ================================================

The main result of this study is the identification of a main-jet axis in the Vela CCSNR. This is the S-shaped structure we draw in Figures \ref{fig:serp} - \ref{fig:overlayed}. We based this identification on the high abundance of ejecta material, namely, O, Ne, and Mg, as the X-ray analysis of Vela by \cite{Mayeretal2023} reveals (Figure \ref{fig:serp}) and on the boundary of the X-ray bright inner zone that we draw on Figures \ref{fig:traced} and \ref{fig:overlayed}. 

In earlier studies (\citealt{Soker2023SNRclass, Soker2024CF}), we discussed the point-symmetric morphology of the Vela CCSNR and its formation in the JJEM based only on outer ears and clumps. Identifying the S-shaped ejecta-rich main-jet axis has two critical implications. (1) The high abundance of O, Ne, and Mg implies that the S-shaped material was ejected during the explosion, as these metals come from the deep core. (2) The large volume that the precessing jets that we take to have formed the S-shaped main-jet axis influenced (the boundary drawn on Figures \ref{fig:traced} and \ref{fig:overlayed}), implies that these jets were very energetic. \cite{GrichenerSoker2017ears}, who considered only the pair DE, found the energy in the jets that inflated these two ears to add up to only $\approx 1 \%$ of the Vela explosion energy. Now, with the other pairs and the two precessing jets that shaped the main-jet axis, the energy adds up to a much more significant fraction of the explosion energy, $\approx 20 \%$. This is compatible with the JJEM, as early jets in the explosion process that supplied the rest of the explosion energy did not leave marks in the morphology.   

As discussed in Section \ref{sec:PointSymmetry}, other shaping processes can not explain the extension of the main-jet axis through the center, the ejecta abundance pattern, the large volume of the main-jet axis, and the point-symmetric wind-rose morphology (see Table \ref{Tab:Table1}). 

We consider the point-symmetric morphologies of CCSNRs to pose the most severe challenge to the neutrino-driven explosion mechanism, as Table \ref{Tab:Table1} shows. It might be that these point-symmetric CCSNRs even rule out the neutrino-driven explosion mechanism. We emphasize that neutrino heating does take place, but in boosting the energy of the jittering jets at launching and in their interaction with the inner core, rather than being the primary explosion process \citep{Soker2022nu}. Studies have identified point-symmetric morphology in about twelve CCSNRs, some with clear point-symmetric wind-rose and some with more subtle point-symmetric morphologies; we expect to increase this number in 2025.

% ===================================================
\section*{Acknowledgements}
% ===================================================
{{{{ We thank an anonymous referee for helpful suggestions. }}}} A grant from the Pazy Foundation supported this research.

\end{document}